\documentclass[preprint,aps,showpacs,amsfonts,epsf]{revtex4}

\input{epsf.tex}

\newcommand{\s}{\sigma}

\renewcommand{\d}{{\rm d}}
\renewcommand{\a}{\alpha}

\newcommand{\be}{\begin{equation}}
\newcommand{\ee}{\end{equation}}
\newcommand{\bea}{\begin{eqnarray}}
\newcommand{\eea}{\end{eqnarray}}
\newcommand{\ba}{\begin{array}}
\newcommand{\ea}{\end{array}}
\def\J#1#2#3#4{{#1} {\bf #2}, #3 (#4)}
\def\PRD{Phys. Rev. D}
\def\PR{Phys. Rev.}

\def\APL{Ann. Phys. (Leipzig)}

\def\CQG{Class. Quantum Grav.}

\def\PLA{Phys. Lett. A}

\def\NCB{Nuovo Chim. B}

\begin{document}
\draft
\title{On the equilibrium of two oppositely charged masses\\ in general relativity}

\author{V.~S.~Manko\,$^\dag$ and E.~Ruiz$\,^\ddag$}
\address{$^\dag$Departamento de F\'\i sica, Centro de Investigaci\'on y de
Estudios Avanzados del IPN, A.P. 14-740, 07000 M\'exico D.F.,
Mexico\\$^\ddag$Instituto Universitario de F\'{i}sica
Fundamental y Matem\'aticas, Universidad de Salamanca, 37008 Salamanca, Spain}

\begin{abstract}
It is shown that the equilibrium configurations of two spherical masses carrying charges of opposite sign that are present in Bonnor's approximate model for two charged particles and also in the exact double--Reissner--Nordstr\"om solution are formed due to the additional electric field of the {\it repulsive} character induced in the black--hole constituent by its overcharged partner. A three--parameter family of such configurations is pointed out. \end{abstract}


\maketitle


A surprising discovery was made in 1993 by Bonnor \cite{Bon} while considering the condition of equilibrium for a spherical charged test particle in the Reissner--Nordstr\"om field \cite{Rei,Nor}: he found that some of the equilibrium configurations allowed the charges to have opposite signs. Obtained with the aid of an approximation scheme, this result probably did not look quite credible, as it was not widely discussed in the literature; besides, Bonnor did not offer a convincing explanation of his effect, restricting himself to noting that ``the gravitational repulsion caused by the charge ... is so strong that the gravitational attraction of the masses alone cannot balance it; electrostatic attraction is needed as well'', thus suggesting the extension of the role of the charge beyond the usual electrostatic interaction.

Nevertheless, the equilibrium of two oppositely charged masses has been recently confirmed within the framework of an exact solution of the static Einstein--Maxwell equations, and one can easily verify by direct substitution into the double--Reissner--Nordstr\"om solution~\cite{Man} that, e.g., the {\it exact} values
$$
M_1=1, \quad Q_1=2, \quad M_2=\frac{1}{5}, \quad Q_2=-\frac{1}{4}, \quad R=\frac{13\sqrt{905}-87}{560}, $$
where $M_l$ and $Q_l$ are, respectively, the Komar masses and Komar charges \cite{Kom} of the sources, and $R$ the separation distance, define a genuine equilibrium configuration between a black hole and a hyperextreme object. So what is the nature of such unusual equilibrium states?

To find an answer to the above question, let us first consider, as a gedankenexperiment, a Schwarzschild black hole placed near a Reissner--Nordstr\"om black hole. Apparently, the electric charge of the latter must induce some electrostatic field in the former black hole which could show up for example as a non--zero value of the electric potential on the Schwarzschild horizon. Then the induced electric field (the net charge of the Schwarzschild black hole being always zero) will interact with the charge of the Reissner--Nordstr\"om black hole, giving rise to a non--zero electrostatic force. In case the latter force is repulsive, it must grow up with growing charge. So, if the induced electric field does not (as will be seen later) produce attraction, we will have a reliable mechanism of electrostatic repulsion that could be applied as well to configurations comprised of two Reissner--Nordstr\"om constituents.

It should be mentioned that the notions of the induced electric field and induced charge on the black--hole horizon were first introduces by Hanni and Ruffini~\cite{HRu} who also studied various properties of the Schwarzschild black hole perturbed by a charged particle. A more sophisticated analysis of the electrostatic two--body problem has been recently performed by Binni, Geralico and Ruffini~\cite{BGR1,BGR2,BGR3} in the framework of the perturbation theory; however, they did not tackle the case of oppositely charged masses, neither their approach permitted them to discover the repulsive character of the force associated with the induced electric field, the result reported in the present letter. Our own analysis of the equilibrium problem and of the associated balance mechanism relies on the exact solution of the field equations and on the use of exclusively analytical formulas describing the interaction of charged masses, so we shall be able to draw principal physical conclusions without resorting to complicated calculations.

In view of the importance of the two--body configurations consisting of one charged and another uncharged black holes for demonstrating in a pure form the presence of the electrostatic repulsive interaction due to induced electric field, let us describe these in more detail. Mathematically, such configurations arise as a four--parameter family of spacetimes obtainable from the general double--Reissner--Nordstr\"om solution by setting to zero one of the Komar charges, $Q_1$ or $Q_2$. Therefore, for the analysis of the `Schwarzschild--Reissner-Nordstr\"om' interaction we can use the general expression for the interaction force between two Reissner--Nordstr\"om constituents obtained in \cite{Man} and having the form
\be {\cal F}=\frac{M_1M_2-(Q_1-\mu)(Q_2+\mu)}{R^2-(M_1+M_2)^2+(Q_1+Q_2)^2}, \quad \mu:=\frac{M_2Q_1-M_1Q_2}{R+M_1+M_2}. \label{force} \ee
Suppose that the upper constituent is the Schwarzschild black hole, then its mass is $M_2$, while the charge $Q_2$ is equal to zero (see Fig.~1a). Setting $Q_2=0$ in (\ref{force}), we get
\be {\cal F}_{\rm Sch-RN}=\frac{M_1M_2-M_2Q_1^2(R+M_1)(R+M_1+M_2)^{-2}}{R^2-(M_1+M_2)^2+Q_1^2}, \label{F4} \ee
whence it follows, taking into account the positiveness of the denominator in (\ref{F4}) for separated constituents, that the second term in the numerator of ${\cal F}_{\rm Sch-RN}$ is a negative quantity independently of the sign of the charge $Q_1$ that carry the Reissner--Nordstr\"om constituent, and hence determines the induced electric force which always has a {\it repulsive} character. That this force cannot lead to attraction can be also seen by considering the values $\Phi_1$ and $\Phi_2$ of the solution's electric potential on the horizons of the two black holes; thus, using for example formulae (8) of \cite{MRS}, we have
\be \Phi_1=\frac{\a Q_1}{M_1+\sqrt{M_1^2-\a Q_1^2}}, \quad
\Phi_2=\frac{Q_1}{R+M_1+M_2}, \quad \a:=\frac{R+M_1-M_2}{R+M_1+M_2}, \label{EP} \ee
so that $\Phi_2$ has the same sign as $\Phi_1$, and hence the induced electric force is indeed repelling. In fact, a distant observer would see the Schwarzschild black hole as a charged black hole with ``charge'' proportional to $Q_1M_2/R$.

Below we write down the metric and electric potential $\Phi$ describing the `Schwarzschild--Reissner-Nordstr\"om' configurations:
$$
\d s^2=f^{-1}[e^{2\gamma}(\d\rho^2+\d z^2)+\rho^2\d\varphi^2]-f\d
t^2, \vspace*{0.5cm} $$ \vspace{-1.5cm} \bea f&=&\frac{A^2-B^2+Q_1^2C^2}{(A+B)^2}, \quad e^{2\gamma}=\frac{A^2-B^2+Q_1^2C^2} {K_0^2 r_1r_2r_3r_4}, \quad \Phi=\frac{Q_1 C}{A+B}, \nonumber \\
A&=&M_2\s_1[\nu(r_1+r_2)(r_3+r_4)+4\kappa(r_1r_2+r_3r_4)] -(Q_1^2\mu^2\nu-2\kappa^2)(r_1-r_2)(r_3-r_4), \nonumber\\ B&=&2M_2\s_1[(M_1\nu +2M_2\kappa)(r_1+r_2)+(M_2\nu+2M_1\kappa)(r_3+r_4)] \nonumber\\ &-&2\s_1[Q_1^2\nu\mu^2+2\kappa(RM_2+Q_1^2\mu-Q_1^2\mu^2)](r_1-r_2) \nonumber\\ &-&2M_2[Q_1^2\nu\mu(1-\mu)-2\kappa(RM_1-Q_1^2\mu^2)](r_3-r_4), \nonumber\\  C&=&2M_2\s_1[(\nu-\nu\mu+2\kappa\mu)(r_1+r_2)+(\nu\mu+2\kappa-2\kappa\mu)(r_3+r_4)] \nonumber\\ &-&2\s_1[M_2\mu\nu+2\kappa(M_1\mu+R\mu)](r_1-r_2) \nonumber\\ &-&2M_2[M_1\mu\nu+2\kappa(M_2\mu+R\mu-R)](r_3-r_4), \label{metric} \eea where
\bea K_0&=&4M_2\s_1[R^2-(M_1-M_2)^2+Q_1^2(1-2\mu)^2], \nonumber \\ \kappa&=&M_1M_2-Q_1^2\mu(1-\mu), \quad \nu=R^2-\s_1^2-M_2^2+2Q_1^2\mu^2, \nonumber \\ \s_1&=&\sqrt{M_1^2-Q_1^2(1-2\mu)}, \quad \mu:=\frac{M_2}{R+M_1+M_2}, \label{not} \eea
and the functions $r_n$ have the form \bea r_1&=&\sqrt{\rho^2+(z-{\textstyle\frac{1}{2}}R-M_2)^2}, \quad r_2=\sqrt{\rho^2+(z-{\textstyle\frac{1}{2}}R+M_2)^2}, \nonumber \\ r_3&=&\sqrt{\rho^2+(z+{\textstyle\frac{1}{2}}R-\s_1)^2}, \quad r_4=\sqrt{\rho^2+(z+{\textstyle\frac{1}{2}}R+\s_1)^2}. \label{rn} \eea

It is worth noting that within the framework of the two--black--holes model it is impossible to achieve balance (i.e., ${\cal F}_{\rm Sch-RN}=0$) of the gravitational attractive and electrostatically induced repulsive forces, even though the latter force is an increasing function of $Q_1^2$. It is easy to see from (\ref{not}) that the Reissner--Nordstr\"om constituent is a charged black hole only when its Komar charge $Q_1$ satisfies the inequality $0<Q_1^2\le M_1^2\a^{-1}$, but even at the maximal value $Q_1^2=M_1^2\a^{-1}$ the corresponding value of ${\cal F}_{\rm Sch-RN}$ is still a positive quantity. That is why the black--hole constituents are prevented from falling onto each other by a supporting strut between them \cite{Isr} representing a conical singularity. Further, when $Q_1^2$ is greater than $M_1^2\a^{-1}$, the Reissner--Nordstr\"om constituent becomes a hyperextreme object and develops a naked singularity. Then, when $M_1^2\a^{-1}<Q_1^2<M_1(R+M_1+M_2)^2(R+M_1)^{-1}$, the gravitational force is still greater than the absolute value of the repulsive electric force, and the strut plays the same role as in the two--black--holes case. However, once the value $Q_1^2=M_1(R+M_1+M_2)^2(R+M_1)^{-1}$ is achieved, we have the balance of forces (${\cal F}_{\rm Sch-RN}=0$) and consequently the equilibrium configuration of the Schwarzschild black hole with a hyperextreme Reissner--Nordstr\"om constituent without a strut emerges (the corresponding exact solution was formally presented in \cite{ABe} without a discussion of the balance mechanism~\cite{Bon2}). For all $Q_1^2$ exceeding the above equilibrium value the induced electric force prevails the gravitational attraction, and a strut is needed already to prevent the constituents from moving away from each other. In this last case, which is also covered by formulas (\ref{metric})--(\ref{rn}) with $\s_1^2<0$, the electrostatic force can take on arbitrarily large negative values.

After having clarified the repulsive character of the induced electric field in a black--hole constituent, we can pass on to the case of two balanced oppositely charged masses. In what follows, we can restrict consideration to only the `subextreme--hyperextreme' configurations (see Fig.~1b) in which the exotic  equilibrium of our interest occurs. We think it is obvious that the main physical picture of the induction effect can not be seriously altered if a black hole, which we earlier assumed to be electrically neutral, carries now a small charge of any sign. In this slightly new situation there appears a net electrostatic force which is sum of two contributions: the one that follows from the usual interaction of the charges $Q_1$ and $Q_2$, and another arising from the electric field induced in the black--hole constituent by the hyperextreme one. If the charge $Q_2$ of the black hole is relatively small in absolute value and its sign is opposite to the sign of the charge $Q_1$ of the hyperextreme object, it is clear that the net electrostatic force can be adjusted (through an appropriate choice of $Q_1$) in such a way that it will balance the gravitational attraction. The balance condition is the equation ${\cal F}=0$, ${\cal F}$ being the interaction force (\ref{force}) of the general double--Reissner--Nordstr\"om solution \cite{Man}, and we have been able to find the following three--parameter family of equilibrium configurations permitting one to describe in a simple way those configurations in which the charges have opposite signs:
\bea M_1&=&R\cos\lambda, \quad M_2=R\sin\lambda, \nonumber\\ Q_1&=&\epsilon R\sqrt{k}\cos\lambda\,\,{\rm cosec}\left(-\frac{\pi}{4}+\frac{\lambda}{2}\right) \sin\left(-\frac{\pi}{4}+\frac{\lambda+\chi}{2}\right), \nonumber\\ Q_2&=&\epsilon R\sqrt{k}\,{\rm cosec}\,\frac{\lambda}{2}\,\sin\lambda\,\sin\frac{\lambda+\chi}{2}, \nonumber\\ k&:=&{\rm cosec}\,(2\lambda+\chi)\sin 2\lambda, \quad \epsilon:=\pm1, \label{fam}  \eea
where $R$, $\lambda$ and $\chi$ are arbitrary constants.

To obtain a particular configuration of the Bonnor type with the aid of (\ref{fam}), one only needs to assign values to $R$, $0<R<\infty$, and to $\lambda$, $0<\lambda<\pi/2$ (in order to have positive masses), and then choose the remaining value of $\chi$ according to the inequality $\chi_0<\chi<-\lambda$, where $\chi_0$, $-2\lambda<\chi_0<-\lambda$, is a solution of the equation
\be \tan\chi=-2\tan\lambda.  \label{x0} \ee
The equilibrium configurations thus obtained will have $Q_1>0$, $Q_2<0$ (for $\epsilon=+1$), and $Q_1<0$, $Q_2>0$ (for $\epsilon=-1$), the Komar quantities $M_2$ and $Q_2$ referring to the black--hole constituent.

Let us mention in conclusion that the presence of a black--hole constituent seems to be the main factor ensuring formation of an equilibrium configuration of charged masses via the induced electric field mechanism, and we are not sure such a mechanism could play a similar role in the systems composed exclusively of hyperextreme objects. Interestingly, the electric field is not induced in the Majumdar--Papapetrou solution \cite{Maj,Pap}, as was already observed in \cite{BGR2}. At the same time, we emphasize that the electric induction does take place in all other extreme black holes which are present in the general double--Reissner--Nordstr\"om solution \cite{Man} and which can carry charges exceeding the masses in absolute value \cite{Emp}. The approximation scheme employed in \cite{BGR2} does not distinguish between different extreme charged black holes and treats all of them as if they were of the Majumdar--Papapetrou type ($|Q|=M$), which is not true in the exact approach.

\section*{Acknowledgments}

We are thankful to Reinhard Meinel for useful correspondence. This work was supported by Project FIS2006-05319 from Ministerio de Ciencia y Tecnolog\'\i a, Spain, and by the Junta de Castilla y Le\'on under the ``Programa de Financiaci\'on de la Actividad Investigadora del Grupo de Excelencia GR-234'', Spain.

\vspace{4cm}

\begin{figure}[htb]
\centerline{\epsfysize=90mm\epsffile{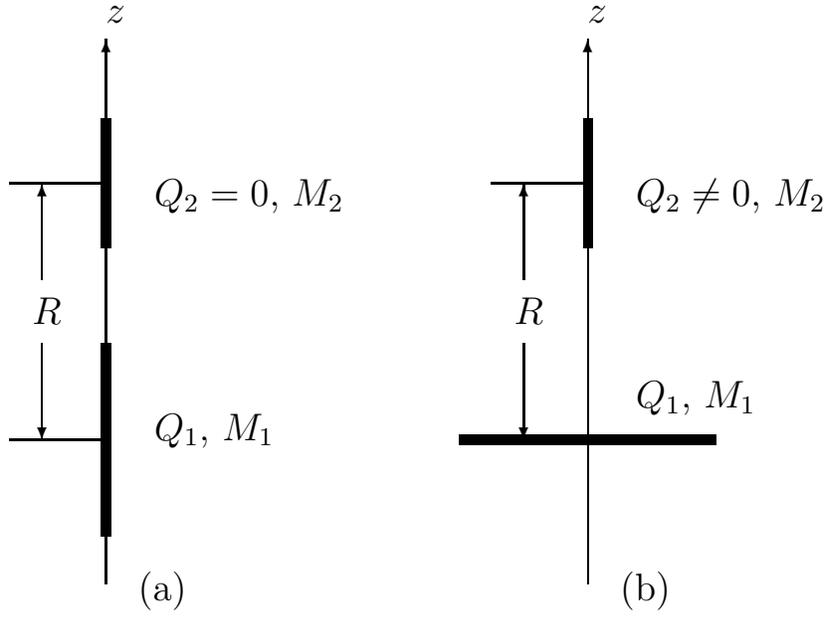}} \caption{Location of sources on the symmetry axis: (a) configuration consisting of a charged and uncharged black holes; (b) configuration consisting of a black--hole and hyperextreme Reissner--Nordstr\"om constituents.}
\end{figure}


\begin{references}
\bibitem{Bon2} In the paper \cite{Bon} the equilibrium configuration of a zero--charge test particle was omitted by Bonnor, but it arises as the solution $q=0$, $r=Q^2/M$ to his equation (9).
\end{references}
\end{document}